\documentclass[11pt,fleqn,twoside]{article}
\usepackage{amsfonts,amssymb,latexsym}
\makeatletter
\newcommand{\prava}[1]{\small\it
\begin{flushleft}
Copyright \copyright \ 1999 by  #1
\end{flushleft}}

\newcommand{\name}[1]{\begin{flushleft}
                       \LARGE \bf #1
                       \end{flushleft}\vspace{-3mm}}

\newcommand{\Author}[1]{\begin{flushleft}
                       \it #1 \end{flushleft}}

\newcommand{\Adress}[1]{\begin{flushleft}
                       \it #1 \end{flushleft}}

\newcommand{\Date}[1]{\begin{flushleft}
                      \small  \it #1 \end{flushleft}}

\newcommand{\ehkol}{Author \ name}
\newcommand{\ohkol}{Article \ name}
\renewcommand{\@evenhead}{
\hspace*{-3pt}\raisebox{-15pt}[\headheight][0pt]{\vbox{\hbox to \textwidth 
{\thepage \hfil \ehkol}\vskip4pt \hrule}}}
\renewcommand{\@oddhead}{
\hspace*{-3pt}\raisebox{-15pt}[\headheight][0pt]{\vbox{\hbox to \textwidth 
{\ohkol \hfil \thepage}\vskip4pt\hrule}}}
\renewcommand{\@evenfoot}{}
\renewcommand{\@oddfoot}{}

\newcommand{\be}{\begin{equation}}
\newcommand{\ee}{\end{equation}}
\newcommand{\ba}{\hspace*{-5pt}\begin{array}}
\newcommand{\ea}{\end{array}}
\newcommand{\p}{\partial}
\newcommand{\ds}{\displaystyle}
\makeatother

\begin{document}
\thispagestyle{empty}
\setcounter{page}{198}
\renewcommand{\ehkol}{D. Blackmore, R. Samulyak and A. Rosato}
\renewcommand{\ohkol}{New Mathematical Models for Particle Flow Dynamics}

\begin{flushleft}
\footnotesize \sf
Journal of Nonlinear Mathematical Physics \qquad 1999, V.6, N~2,
\pageref{blackmore-fp}--\pageref{blackmore-lp}.
\hfill {\sc Article}
\end{flushleft}

\vspace{-5mm}

\renewcommand{\footnoterule}{}
{\renewcommand{\thefootnote}{}
 \footnote{\prava{D. Blackmore, R. Samulyak and A. Rosato}}}

\name{New Mathematical Models \\ for Particle Flow Dynamics}\label{blackmore-fp}

\Author{Denis BLACKMORE~$^\dag$, Roman SAMULYAK~$^\dag$ and
Anthony ROSATO~$^\ddag$}

\Adress{$\dag$~Department of Mathematical Sciences, New Jersey Institute of Technology,\\
~~Newark, New Jersey 07102 -- 1982, USA\\
~~E-mail: deblac@chaos.njit.edu, \  rosamu@eclipse.njit.edu\\[2mm]
$\ddag$~Department of Mechanical Engineering and Particle Technology Center,\\
~~New Jersey Institute of Technology,  Newark, New Jersey 07102 -- 1982, USA\\
~~E-mail: rosato@megahertz.njit.edu}

\Date{Received October 16, 1998; Revised December 28, 1998;
Accepted January 8, 1999}

\begin{abstract}
\noindent
A new class of integro-partial dif\/ferential equation
models is derived for the prediction of granular f\/low dynamics. These models
are obtained using a novel limiting averaging method (inspired by techniques
employed in the derivation of inf\/inite-dimensional dynamical systems models) on
the Newtonian equations of motion of a many-particle system incorporating
widely
used inelastic particle-particle force formulas. By using Taylor series
expansions,
these models can be approximated by a system of partial dif\/ferential equations
of the Navier-Stokes type. The exact or approximate governing equations
obtained
are far from simple, but they are less complicated than most of the continuum
models now
being used to predict particle f\/low behavior. Solutions of the new models for
granular
f\/lows down inclined planes and in vibrating beds are compared with known
experimental
and analytical results and good agreement is obtained.
\end{abstract}



\section{Introduction}

The last two decades have witnessed an intensif\/ication of research in
granular f\/low dynamics, in large measure spurred by a burgeoning array of
engineering and industrial applications of particle technology. There are
several features that make granular f\/low research attractive to engineers,
mathematicians and scientists, among which are the following: A need still
exists to formulate the underlying principles of particle interactions in a
completely satisfactory manner; there are as yet few if any def\/initive
mathematical models that can reliably predict a wide range of granular
f\/lows; particle f\/low phenomena such as arching, surface waves and convection
are still not entirely understood from a mathematical or engineering
perspective; there is a panoply of extremely complex nonlinear dynamical
behaviors exhibited in granular f\/low regimes that has not yet been fully
analyzed and has severely tested or exceeded the capabilities of current
experimental and computer technologies for accurate characterization; and
the techniques and devices for optimizing certain features of particle f\/lows
are for the most part only understood on an ad hoc basis. In this paper we
use some averaging and limiting ideas associated with inf\/inite-dimensional
dynamical systems theory to derive a new class of continuum mathematical
models for granular f\/low that may be capable of predicting the dynamical
characteristics of particle f\/lows in a large variety of circumstances, and
thereby help to make some progress in solving the many outstanding problems
in this f\/ield. Our purpose is not to compete with the host of interesting
models that try to incorporate as much of the physics of granular f\/lows as
possible, including (vibrational) energy equations. Rather, we aim to
produce mathematical models that are relatively tractable and ignore just
enough of the physics to still provide useful predictions of granular f\/low
dynamics for a wide range of applications.

Although there have been several partial successes in recent
years, the state-of-the-art in mathematical modeling of granular
f\/low phenomena pales in comparison to that of f\/luid mechanics
where there is a universally accepted model -- the Navier-Stokes
equations~-- whose reliability has been tested and conf\/irmed for
over a century, and is considered in many quarters to be capable
of apprehending even what may be the most elusive of all physical
processes~-- f\/luid turbulence. Approaches based on continuum
mechanics (transport theory) and kinetic theory (statistical
mechanics) have been those most often used for obtaining
mathematical models for particle f\/lows in the form of systems of
partial dif\/ferential equations. Notable examples derived using
these methods which have enjoyed some success in predicting
granular f\/low dynamics may be found in An \& Pierce~\cite{blackmore:An},
Anderson \& Jackson~\cite{blackmore:And}, Farrel, Lun \& Savage~\cite{blackmore:Far},
Gardiner \& Schaef\/fer~\cite{blackmore:Gard}, Goldshtein \&
Shapiro~\cite{blackmore:Gold}, Jenike \& Shield~\cite{blackmore:Jeni}, Jenkins \&
Savage~\cite{blackmore:Jen}, Jenkins \& Richman~\cite{blackmore:Jen2}, Johnson, Nott
\&
 Jackson~\cite{blackmore:John}, Lun~\cite{blackmore:Lun}, Lun \& Savage~\cite{blackmore:Lun3},
Numan \& Keller~\cite{blackmore:Num}, Pasquarell~\cite{blackmore:Pasq}, Pitman~\cite{blackmore:Pit},
Rajagopal~\cite{blackmore:Raj}, Richman~\cite{blackmore:Rich}, Tsimring \& Aranson~\cite{blackmore:Tsim},
Savage~\cite{blackmore:Sav,blackmore:Sav2}, Savage \& Jef\/frey~\cite{blackmore:Sav3}, Schaef\/fer~\cite{blackmore:Scha},
Schaef\/fer, Shearer \& Pitman~\cite{blackmore:Scha2} and Shen \& Ackermann~\cite{blackmore:Shen}
(see also Fan \& Zhu~\cite{blackmore:Fan} and Walton~\cite{blackmore:Walt}). Several of these models have
proven to be rather ef\/fective in
characterizing certain granular f\/lows, for example in chutes and hoppers
with simple geometries, but they tend to be fairly complicated systems of
nonlinear partial dif\/ferential equations that are dif\/f\/icult to analyze and
solve except by approximate numerical methods, and the information they
provide has barely made a dent in the host of practical problems associated
with industrial uses of particle technology. There have also been a number
of simple, idealized models formulated by neglecting a variety of physical
factors, but these tend to miss many of the features of granular f\/low of
interest in applications. Much work still remains in f\/inding a really
ef\/fective balance between mathematical tractability and adherence to the
underlying principles of physics in the models for a large class of granular
f\/low phenomena. It is hoped that the models introduced here will provide
a useful step in the direction of achieving such a balance.

Granular f\/lows have also been extensively studied using methods inspired by
molecular dynamics research. The basic idea of the molecular dynamics
approach is to use realistic models for interparticle forces, developed from
both theory and empirical investigations, in a Newtonian dynamics context
with a large number of particles (hundreds or thousands) to determine the
evolution of a particle f\/low conf\/iguration. Analytical means are of little
use in solving the very high dimensional dynamical systems encountered in
such an approach, but some very sophisticated simulations, employing a
variety of numerical solution techniques, have been devised for studying
granular f\/lows, such as those of Goldhirsch et al.~\cite{blackmore:Goldh},
Lan \& Rosato~\cite{blackmore:Lan2,blackmore:Lan},
McNamara \& Luding~\cite{blackmore:McN}, P{\"o}eschel \& Herrmann~\cite{blackmore:Poesch},
 Rosato et al.~\cite{blackmore:Ros4}, Swinney et al.~\cite{blackmore:Swin} and Walton~\cite{blackmore:Walt}.
Alternative approaches based on  cellular automata models and
kinetic models of random walks in discrete lattices have also proven to be
quite useful; see, for example,
 Baxter \& Behringer~\cite{blackmore:Baxter} and Caram \& Hong~\cite{blackmore:Caram}.
These and other
simulations have proven to be so remarkably accurate in manifesting most of
the complex aspects of particle f\/low behavior, that one is inescapably drawn
to the conclusion that the formulation of a more concise and tractable
mathematical representation of such simulations should greatly enhance our
ability to analyze particle f\/low phenomena.

It was this idea of f\/inding more succinct ways of mathematically
characterizing granular f\/low simulations for extremely large numbers of
particles that served as the inspiration for the new models derived in this
paper by computing limiting forms of the relevant Newtonian dynamical
systems. To be more precise, we obtain systems of nonlinear partial
dif\/ferential equations~-- inf\/inite-dimensional dynamical systems~-- for
velocity f\/ields of granular f\/lows by using an averaging method together with
the computation of a limit as the number of particles tends to inf\/inity,
followed by a Taylor series approximation. The approach employed is akin to the
methods used to obtain limiting partial dif\/ferential equations for systems
of ordinary dif\/ferential equations (as the size of systems tend toward
inf\/inity) in the theory of inf\/inite-dimensional dynamical systems; for
example, as when the Korteweg-de Vries equation is obtained as the ''limit''
of an inf\/inite string of coupled nonlinear oscillators (cf. Tabor~\cite{blackmore:Tab}
and Temam~\cite{blackmore:Tem}). Our method leads to an inf\/inite class of mathematical models
of widely varying levels of complexity, depending on the form of the
particle-particle force laws chosen and the order of the Taylor series
expansions employed. Several of these models appear to enjoy certain
advantages over existing models in terms of simplicity and ease of analysis,
and they have the potential for providing a better developed mathematical
understanding of granular f\/low phenomena.

This paper is organized as follows:  The
particle-particle models, based on the Hertz-Mindlin theory and some
empirical observation, that we shall employ for the granular f\/lows under
consideration are described in Section~2. Then, in Section~3 we develop the
Newtonian dif\/ferential equations of motion for the particle f\/low dynamics
using the particle-particle force formulas introduced in Section~2, and we
describe a decomposition of the forces into interparticle forces, body
forces and transmitted forces. In Section~4 we delineate a limiting procedure
on the Newtonian equations of motion of the granular f\/low, ignoring boundary
contributions, that produces a system of integro-partial dif\/ferential
equations that models the velocity f\/ield of the particle f\/low. This class of
mathematical models for particle f\/low dynamics is inf\/inite and depends on
the form and parameters of the particle-particle force formulas. We also
show how our integro-dif\/ferential equation models can yield an inf\/initude of
partial dif\/ferential equation approximations to the governing equations when
the dynamical variables are approximated by Taylor series. A choice of
interparticle parameters and order of Taylor series expansion that leads to
a particularly simple system of partial dif\/ferential equations for the
particle velocities in a granular f\/low is also described in this section. We
compute, in Section~5, an exact solution of the simple model of Section 6
for a fully developed f\/low through a vertical pipe with a uniform circular
cross-section, and in so doing give our f\/irst illustration of how to append
relevant boundary conditions to our system of equations. The model of
Section 4 and the introduction of appropriate auxiliary conditions necessary
to model fully developed, two-dimensional granular f\/low down an inclined
plane is treated in Section~6, and we compare our results with those
obtained from other analytical and experimental studies of inclined plane
f\/lows. In Section~7 we develop the boundary conditions for f\/lows in a
vibrating bed and study numerically our model subject to these boundary
conditions. We discuss our results and compare them with experimental studies.
Finally, we conclude in Section~8 with a discussion of the
consequences of the work in this paper and possible directions for future
research involving the new class of models.

\section{Interparticle forces}

In this section we shall describe the particle-particle force models that
are the foundation upon which we construct our derivation of the governing
equations of motion for the granular f\/lows under consideration. We assume
that the f\/low system is comprised of a large number $N$ of identical
inelastic spherical particles distributed throughout some region in ${\bf R}
^3$ at points ${\bf x}^{(i)}$, $1\leq i\leq N.$ The common radius of all
the particles is a very small positive number that we denote by $r,$ and the
point ${\bf x}^{(i)}$ corresponding to the $i$th particle is located at
the center of the particle for all $1\leq i\leq N.$

We may select any particle, say the $i$th one, and suppose that the $j$th
particle at ${\bf x}^{(j)}$ is near ${\bf x}^{(i)}.$  For convenience, we
def\/ine
\begin{equation}
{{\bf r}}_i^j:={{\bf x}}^{(j)}-{{\bf x}}^{(i)}
\end{equation}
and ${\bf v}_i^j$ to be the velocity of the $j$th particle relative to
the $i$th particle; namely,
\begin{equation}
{{\bf v}}_i^j:={{\bf v}}\left({{\bf x}}^{(j)}\right)-{{\bf v}}\left({{\bf x}}^{(i)}\right).
\end{equation}
Taking our cue from Hertz-Mindlin theory as supported by numerous
experimental observations and granular f\/low simulations 
(see~\cite{blackmore:Fan,blackmore:Jeni,blackmore:Ros4,blackmore:Sav}
and~\cite{blackmore:Walt}), we shall assume that the model for the force ${\bf P}_i^j$
exerted on the $i$th particle by the $j$th particle is described as follows:
${\bf P}_i^j$ is the sum of a (inelastic) {\it normal force} ${\cal N}
_i^j$ and a {\it tangential} {\it force} ${\cal T}_i^j$ due to
friction
\begin{equation}
{{\bf P}}_i^j={\cal N}_i^j+{\cal T}_i^j,
\end{equation}
where
\begin{equation}
{\cal N}_i^j:=\left[ -\chi \left(\left\| {{\bf r}}_i^j\right\| ^2\right)\left\|
{{\bf r}}_i^j\right\| ^{{}\alpha }+\eta \left(\left\| {{\bf r}}
_i^j\right\| ^2\right)\left\langle {{\bf v}}_i^j,{{\bf r}}
_i^j\right\rangle \left\| {{\bf r}}_i^j\right\| ^{{}\beta }\right]
{{\bf \hat{r}}}_i^j
\end{equation}
and
\begin{equation}
{\cal T}_i^j:=\psi \left(\left\| {{\bf r}}_i^j\right\| ^2\right)\left\| {
{\bf r}}_i^j\right\| ^{{}\gamma }\left\| \vartheta ({\bf v}
_i^j)\right\| ^{{}\delta }\widehat{\vartheta ({\bf v}_i^j)}.
\end{equation}
Here $\langle \cdot  , \cdot \rangle$ is the standard inner product with induced
norm $\|\cdot\|$ in
${\bf R}^3$  and
$\alpha$, $\beta$, $\gamma$ and $\delta $ are positive exponents chosen
according to the particular properties of the material particles; among the
most often used values are $\alpha =1$ or  $3/2$ (Hertzian), $\beta =1,$
$\gamma =0,1/3,2/3$ or $3/2$ and $\delta =1$ or $2.$ The functions $\chi$,
$\psi $ and $\eta $ are smooth ($=C^\infty $) on $[0,\infty )$ and have
the following properties:
\begin{equation}
\chi (\tau ),\psi (\tau ),\eta (\tau )\geq 0\qquad \mbox{for all }\quad \tau \geq 0;
\end{equation}

\begin{equation}
\chi ^{\prime }(\tau ),\psi ^{\prime }(\tau ),\eta ^{\prime }(\tau )\leq
0\qquad \mbox{for all }\quad \tau \geq 0\quad (^{\prime }=d/d\tau );
\end{equation}
\begin{equation}
\chi (\tau )=\psi (\tau )=\eta (\tau )=0\qquad \mbox{when }\quad \tau >4r^2;
\end{equation}
\begin{equation}
\chi ^{\prime }(\tau )=\psi ^{\prime }(\tau )=\eta ^{\prime }(\tau )=0\qquad
\mbox{for }\quad 0\leq \tau \leq q^2<4r^2;
\end{equation}
\begin{equation}
\eta (0)<\chi (0);
\end{equation}
and
\begin{equation}
\eta (\tau )\leq \chi (\tau )\qquad \mbox{for all }\quad \tau \geq 0.
\end{equation}
Graphs of these functions are shown in Figure~1.
As we are going to ignore the rotational motion of the particles in our
treatment, we shall assume that the tangential force is very small compared
to the normal force and, more specif\/ically, that $\psi(\tau)\ll \eta(\tau)$
for all $\tau$ such that $\psi(\tau) > 0$.

\strut\hfill

{\bf Figure 1:} Force functions: a) $y=\chi(x)$, b) $y=\eta(x)$, c) $y=\psi(x)$

\vspace{9cm}

The role of the function $\eta $ is to represent an energy loss due to
inelasticity in the restoring mode when the particles are separating after a
collision. A caret over a vector ${\bf u} $ indicates the unit vector in the
direction of ${\bf u}$; i.e., ${\bf \hat{u}}$:=${\bf u}/\left\| {
{\bf u}}\right\| .$ The vector $\vartheta \left({\bf v}_i^j\right)$ is the
component of the relative velocity at the point of contact of a pair of
particles obtained by projecting ${\bf v}_i^j$ onto the tangent plane of
the $i$th particle at the point of contact. This vector can be written in
the form
\begin{equation}
\vartheta \left({{\bf v}}_i^j\right):={{\bf v}}_i^j-\left\langle {
{\bf v}}_i^j,{{\bf \hat{r}}}_i^j\right\rangle {{\bf \hat{r}}}
_i^j.
\end{equation}
The normal and tangential interparticle forces are depicted in Figure~2.

\strut\hfill

{\bf Figure 2:}
Particle-particle forces: a) normal force, b) tangential force

\vspace{10cm}

Summing over all particles in the granular f\/low system, we f\/ind that the
total force exerted by all the particles on the $i$th particle is
\begin{equation}
{{\bf P}}_i:=\sum_{j=1,j\neq i}^N{{\bf P}}
_i^j=\sum_{j=1,j\neq i}^N\left( {\cal N}_i^j+{\cal T}_i^j\right) .
\end{equation}
Note that our assumed particle-particle force models account for the
geometry of the particles only with regard to the region where the force
vanishes (its {\it support)} and the manner in which the tangential
frictional component of force is def\/ined. We observe that~(13) can also be
obtained from a specif\/ic force density f\/ield surrounding the $i$th
particle with the force supplied by each grain equal to this specif\/ic
density multiplied by the volume $\frac 43\pi r^3$. We shall return to this
point in the sequel when we compute limiting forms of the particle dynamical
system.

\section{Newtonian equations of motion}

The motion of the particles in the granular f\/low f\/ield may be described by a
system of $3N$ second-order, ordinary dif\/ferential equations expressing
Newton's second law of motion; viz.
\begin{equation}
m{{\bf \ddot{x}}}^{(i)}={{\bf F}}_i:={{\bf P}}_i+{\bf T}_i+{
{\bf E}}_i+{{\bf B}}_i\qquad (1\leq i\leq N),
\end{equation}
where $\dot{}=d/dt,$ $m$ is the mass of each of the $N$
identical particles, {\bf P}$_i$ is the force exerted
 on the $i$th particle by all particles in direct contact with it
as described in the preceding section,
${\bf T}_i$ is the {\it transmitted force} on the $i$th particle exerted by
connected arrays of particles in contact with one another that touch a
particle in direct contact with the $i$th particle,
{\bf E}$_i$ is the {\it external} or {\it body force} on the $i$th
particle which is usually just the gravitational force (but may sometimes
also include electromagnetic and other forces) and {\bf B}$_i$ is the
{\it boundary} {\it force }exerted on the $i$th particle
 by f\/ixed or motile boundaries in direct contact with it
that delimit the
region in space in which the particles can move. Observe that the variables
on which each of the components of force depend can be described as follows:
\[
{{\bf P}}_i={{\bf P}}_i\left( {{\bf x}}^{(1)},\ldots,
{{\bf x}}^{(N)},{{\bf \dot {x}}}^{(1)},\ldots, {{\bf \dot {x}}}^{(N)}\right),
\]
\[
{{\bf T}}_i={{\bf T}}_i\left( {{\bf x}}^{(1)},\ldots,
{{\bf x}}^{(N)},{{\bf \dot {x}}}^{(1)},\ldots, {{\bf \dot {x}}}
^{(N)}\right),
\]
\[
\frac{{{\bf E}}_i}m\quad  \mbox{is usually constant},
\]
\[
{{\bf B}}_i={{\bf B}}_i\left( {{\bf x}}^{(i)},{{\bf \dot {x}}}^{(i)},t\right) ,
\]
where the dependence on $t$ in ${\bf B}_i$ occurs when the material
boundary of the f\/low region moves with time such as in the case of particles
moving in a vibrating container.

We can write the Newtonian equations of motion in a more concise form by
introducing the following vector notation: Def\/ine the vector ${\bf X}$ in $
{\bf R}^{3N}$ to be
\[
{{\bf X}}:=\left( {{\bf x}}^{(1)},{{\bf x}}^{(2)},\ldots,{{\bf x}}^{(N)}\right) .
\]
Then (14) can be rewritten in vector form as
\begin{equation}
{{\bf \ddot{X}}}={\bf \Phi }({{\bf X}},{{\bf \dot {X}}}
,t):={\bf \Phi }_p({\bf X},{\bf \dot{X}})+{\bf \Phi }_T({\bf X},{\bf \dot{X}},t)+
{\bf \Phi }_e+{\bf \Phi }_b({\bf X},{\bf \dot{X}},t),
\end{equation}
where
\[
{\bf \Phi }_p:=m^{-1}\left( {{\bf P}}_1({{\bf X}},{
{\bf \dot {X}}}),\ldots,{{\bf P}}_N({{\bf X}},{{\bf
\dot {X}}})\right)
\]
is the {\it interparticle force per unit mass,}
\[
{\bf \Phi}_T:=m^{-1}({\bf T}_1({\bf X}, {\bf \dot X}, t),
\ldots, {\bf T}_N({\bf X}, {\bf \dot X}, t))
\]
is the {\it transmitted force per unit mass},
\[
{\bf \Phi }_e:=m^{-1}\left( {{\bf E}}_1,\ldots,{{\bf E}} _N\right)
\]
is the {\it external force per unit mass} and
\[
{\bf \Phi }_b:=m^{-1}\left( {{\bf B}}_1({{\bf x}}^{(1)},
{{\bf \dot {x}}}^{(1)},t),\ldots,{{\bf B}}_N({{\bf x}}
^{(N)},{{\bf \dot {x}}}^{(N)},t)\right)
\]
is the {\it  boundary force per unit mass}. In theory, if initial values
of ${\bf X}$ and ${\bf \dot{X}}$ are specif\/ied, then~(15) uniquely
determines the ensuing motion of all the particles, at least for small
values of $\left| t\right| $ (see~\cite{blackmore:Tab} and~\cite{blackmore:Tem}). However, for
extremely
large values of $N$ the work required to integrate~(15)~-- analytically, when
in the rare cases that this is possible, or numerically otherwise~-- tends to
be prohibitive. Thus it is desirable to f\/ind an inf\/inite-dimensional limit
in some sense for~(15) as $N\rightarrow \infty ,$ presumably in the form of
a partial dif\/ferential equation, that may prove to be more amenable to
analysis. This is precisely what we shall do in the next section.

\section{Limiting models}

We shall demonstrate how new models for granular f\/low phenomena can be
obtained by applying a certain type of dynamical limit procedure to the
Newtonian equations~(15). The reader will no doubt notice at least a vague
similarity between our method and the continuum limit used in the
Fermi-Ulam-Pasta model to obtain the Korteweg-de Vries equation (cf.~\cite{blackmore:Tab}).
To begin with, we restrict our attention to points in the interior of the
granular f\/low region that are not directly af\/fected by interaction with the
boundary. Consequently, for the time being we ignore the boundary force
contribution in~(14) or~(15); the boundary ef\/fects shall be considered in
the sequel when we study specif\/ic boundary-value problems.

Referring to (14), we assume that the body forces are exclusively
gravitational and that the Cartesian coordinate system has been chosen so
that the gravitational force acting on each particle has the form
\begin{equation}
{\bf E}_i=-m g{\bf \hat{e}},
\end{equation}
where $g$ is the acceleration of gravity and ${\bf \hat{e}}$ is a unit
vector in the opposite direction to the gravitational f\/ield. Now we
select a point in the interior of the granular f\/low f\/ield corresponding to
the $i$th particle (at time~$t$) which is moving along a trajectory
determined by the vector f\/ield
\begin{equation}
{\bf \dot{x}}^{(i)}={\bf v}\left( {\bf x}^{(i)},t\right)
\end{equation}
and the location of this particle at time $t=0.$

The interparticle forces on the $i$th particle at the point ${\bf x}^{(i)}={\bf x}$
are given by~(13). Since we are going to take a limit as
the number of grains goes to inf\/inity, we need to average or distribute
these forces in a way that insures the existence of such a limit and is
conducive to its computation. This can be done by smearing the particles
into a continuum (assumed to be locally uniform)
and considering the interparticle force f\/ield to be
obtained from a specif\/ic force density f\/ield. To be more precise, we assume
that each particle is surrounded by a specif\/ic force density f\/ield of the
same form $c{\bf P}_{*}$. Whence, the interparticle
force on the $i$th particle can be written
\begin{equation}
{\bf P}_i:=c\sum_{j\neq i}{{\bf P}}_{*}\left({{\bf x}}^{(j)};{\bf v}_i^j\right)\Delta V_j,
\end{equation}
where $\Delta V_j$ is the volume increment occupied by the $j$th particle
and $c>0$ is a multiplicative factor with units {\it volume}$^{-1}$
(associated with the geometry of the particles). This can be rewritten in
the form
\begin{equation}
{\bf P}_i=(N-1)^{-1}cc_0\sum_{j\neq i}{\bf P}_{*}\left({\bf x} ^{(j)};{\bf v}_i^j\right),
\end{equation}
where $c_0$ is a positive constant equal to the volume of the (compact)
support of ${\bf P}_i$ and we have assumed that all the particles occupy
volume increments of the same size. In~(19) we plainly see the averaging
aspect of this approach. Taking the limit as $N\rightarrow \infty $ in~(19)
[or equivalently as $\Delta V_j$ $\rightarrow 0$ in~(18)] using standard
results from integration theory, we obtain
\begin{equation}
\lim_{N\rightarrow \infty }{\bf P}_i=c\int_{{\bf R}^3}{\bf P}
_{*}\, dy_1dy_2dy_3=c\int_{{\bf R}^3}{\bf P}_{*}\, d{\bf y},
\end{equation}
where it follows from (3), (4), (5) and (12) that
\be
\ba{l}
{\bf P}_{*} :={\bf P}_{*}\left( {\bf y;v(x),v}({\bf x}+{\bf
y})\right)
\vspace{2mm}\\
\ds \qquad = \left[ -\chi \left(\left\| {\bf y}\right\| ^2\right)\left\| {\bf y}
\right\| ^{\alpha -1}+\eta \left(\left\| {\bf y}\right\| ^2\right)\left\langle
{\bf v}({\bf x}+{\bf y})-{\bf v}({\bf x}),{\bf y}
\right\rangle \left\| {\bf y}\right\| ^{\beta -1}\right] {\bf y}
\vspace{2mm}\\
\ds \qquad +\psi \left(\left\| {\bf y}\right\| ^2\right)\left\| {\bf y}\right\| ^{{}\gamma
}\left\| {\bf v}({\bf x}+{\bf y})-{\bf v}({\bf x}
)-\left\langle {\bf v}({\bf x}+{\bf y})-{\bf v}({\bf x}),
{\bf y}\right\rangle \frac{{\bf y}}{\left\| {\bf y}\right\| ^2}
\right\| ^{\delta -1}
\vspace{2mm}\\
\ds \qquad \times \left[ {\bf v}({\bf x}+{\bf y})-{\bf v}({\bf x}
)-\left\langle {\bf v}({\bf x}+{\bf y})-{\bf v}({\bf x}),
{\bf y}\right\rangle \frac{{\bf y}}{\left\| {\bf y}\right\| ^2}
\right] ,
\ea
\ee
where ${\bf y}=(y_1,y_2,y_3)$ represents the position vector measured
from the reference point~${\bf x}$ that has been introduced to simplify
the notation for ${\bf r}_i^j.$
As for the transmitted force in the Newtonian equation~(14), we make the
standard assumption that in the continuum limit it can be represented by
a gradient f\/ield, $\mbox{grad}\,p$, where the function $p=p({\bf x},t)$ is
naturally called the {\it pressure}. The particle at ${\bf x}^{(i)}$ is
represented by a {\it density  field}, $\rho=\rho({\bf x},t)$, with compact support. Hence
the external force is
\[
\left(\int_{W_i}\rho g\,dV\right){\bf \hat e},
\]
where $W_i$ is a spherical (control) region centered at ${\bf x}^{(i)}$ with
(Lebesgue) measure $\Delta V_i$, and this converges to $\rho g{\bf \hat e}$ as
$N\to\infty$ ($\Leftrightarrow \Delta V_i\to 0$). In the same spirit, the right-hand side of~(14)
is replaced by
\[
\frac d{dt}\int_{W_i}\rho {\bf v}\,dV,
\]
which upon applying the usual continuum limit converges to the density times
the total (material) derivative of the velocity:
\[
\rho\frac{D{\bf v}}{ Dt}:=\rho\left(\frac {\p{\bf v}}{\p t}+
\sum^3_{k=1}\frac {\p{\bf v}}{\p x_k}v_k\right).
\]
Upon combining all of the above computations, we obtain the following system of
nonlinear integro-partial dif\/ferential equations for the momentum balance of
the particle f\/low in the interior of the region under consideration:
\be
\frac{D {\bf v}}{Dt}=\frac {\p {\bf v}}{\p t}+v_k\frac {\p {\bf v}}{\p x_k}=
-g{\bf \hat e}+ \frac 1\rho\mbox{grad}\,p+\kappa\int_{{\bf R}^3}{\bf P}_*({\bf y};
{\bf v}({\bf x}), {\bf v}({\bf x}+{\bf y}))\, dy,
\ee
where $\kappa:=c/\rho$ and we have employed the Einstein summation convention.
If $\rho$ is constant and $\mbox{grad}\,p$ is known a priori, then~(22) together
with appropriate initial and boundary data suf\/f\/ices to determine the
velocity f\/ield. When the granular f\/low is compressible and $\mbox{grad}\,p$ is
known a priori, we have to add the continuity equation
\be
\frac{\p\rho}{\p t}+\mbox{div}\,(\rho{\bf v})=0
\ee
to (22), and then by imposing additional auxiliary data the velocity f\/ield and
density may be determined. Of course, in general, both~$p$ and~$\rho$ are unknown
variables, in which case~(22) and~(23) are insuf\/f\/icient to determine~$p$, $\rho$ and
${\bf v}$. One more equation must be added, and this may be accomplished by appending an
energy equation to~(22) and~(23). The easiest way to do this is to obtain an equation
of state of the granular f\/low medium that provides a relationship between the
pressure and the density. If we make the same assumption as above (in particular, that
the particles are uniformly and isotropically distributed locally), then by applying the same
type of limit as $N\to\infty$ to the equations representing the kinetic energy
of the Newtonian system~(14), we obtain the equation of state of an ideal gas; namely
\be
p=A\,\rho^\omega,
\ee
where $A>0$ and $\omega>1$ are constants that are obtained from the properties of
the granular f\/low medium. The same result can be derived by applying the standard
thermodynamic limit of statistical mechanics in conjunction with the virial theorem.

For certain purposes, including comparison with other continuum models for
particle f\/lows, it is useful to replace~(22) with an approximate partial dif\/ferential equation.
Although, it should be pointed out that, mathematically speaking,~(22) enjoys
certain inherent advantages over such partial dif\/ferential equation models. In
particular, as we shall demonstrate in a forthcoming paper, solutions of the system with~(22)
exhibit considerably more regularity than the pure dif\/ferential equation models
that we shall discuss in the sequel.

In order to approximate (22) by a system of $m$th order partial dif\/ferential
equations, we may use the following Taylor series expansion of order $m:$
\begin{equation}
{\bf v}({\bf x}+{\bf y})-{\bf v}({\bf x})\simeq
\sum_{k=1}^m\frac 1{k!}\frac{\partial ^k{\bf v}}{\partial {\bf
x}^k}({\bf x}){\bf y}^k.
\end{equation}
The positive integer $m$ is at our disposal, and it is plausible to assume
that the larger we choose~$m$, the more accurate the approximation.
Substituting~(25) in~(22), we obtain the system of $m$th order, nonlinear
partial dif\/ferential equations as an approximate model for the momentum balance
of the granular f\/low f\/ield given by
\begin{equation}
\frac{\partial {\bf v}}{\partial t}+v_k\frac{\partial {\bf v}}{
\partial x_k}=-g{\bf \hat e}+\frac 1\rho \, \mbox{grad}\,p+{\bf \Gamma }\left( {\bf
v},\frac{\partial {\bf v}}{\partial {\bf x}},\ldots,\frac{\partial ^m{\bf v}}{\partial
{\bf x}^m}\right) ,
\end{equation}
where{\bf \ }${\bf \Gamma },$ a function that does not depend
explicitly on ${\bf x},$ is def\/ined by
\[
\ba{l}
\ds {\bf \Gamma } :=\kappa \left\{ \!\! -\int_{{\bf R}^3} \!\chi \left(\left\|
{\bf y}\right\| ^2\right)\left\| {\bf y}\right\| ^{\alpha -1}{\bf y}d
{\bf y}+ \! \sum_{k=1}^m\frac 1{k!}\int_{{\bf R} ^3} \! \left\langle \frac{\partial ^k{\bf v}}{\partial 
{\bf x}^k}{\bf y}
^k,{\bf y}\right\rangle \left\| {\bf y}\right\| ^{\beta -1}\eta
\left(\left\| {\bf y}\right\| ^2\right){\bf y}d{\bf y}\right.
\vspace{3mm}\\
\ds \qquad +\sum_{k=1}^m\frac 1{k!}\int_{{\bf R}^3}\psi \left(\left\|
{\bf y}\right\| ^2\right)\left\| {\bf y}\right\| ^{{}\gamma }\left\|
\sum_{k=1}^m\frac 1{k!}\left[ \frac{\partial ^k{\bf v}}{\partial
{\bf x}^k}{\bf y}^k-\left\langle \frac{\partial ^k{\bf v}}{\partial
{\bf x}^k}{\bf y}^k,{\bf y}\right\rangle \frac{{\bf y}}{\left\|
{\bf y}\right\| ^2}\right] \right\| ^{\delta -1}
\vspace{3mm}\\
\ds \qquad \times \left. \left[ \frac{\partial ^k{\bf v}}{\partial {\bf x}^k}{\bf y}
^k-\left\langle \frac{\partial ^k{\bf v}}{\partial {\bf x}^k}{\bf y}
^k,{\bf y}\right\rangle \frac{{\bf y}}{\left\| {\bf y}\right\| ^2}
\right] d{\bf y}\right\} .
\ea
\]
Hence we have inf\/initely many possible partial dif\/ferential equation models
for granular f\/low corresponding to the choices of the functions $\chi$, $\eta $
and $\psi$, of parameters $\alpha$, $\beta$, $\gamma $ and $\delta$, and the
order~$m$ of the Taylor series approximation.
This leads to a very natural question: What order of Taylor series
approximation in~(26) should be used for a given application? As we shall show in the sequel,
$m=2$ works rather well for tube, inclined plane and vibrating bed f\/lows.
However, it will probably be necessary to consider several choices in other
applications and determine an acceptable order of approximation on a
case-by-case basis, where an educated guess is made based upon known properties
of the f\/low.

\subsection*{A simple f\/low model}

Depending on the choice of parameters and the order, the equation~(26) can
range from relatively simple to quite complicated. In this section we make a
choice of parameters and order that leads to a rather simple yet ostensibly
realistic model for the velocity f\/ield of a granular f\/low. Specif\/ically, we
choose $\alpha =\beta =1,$ $\gamma =0$, $\delta =1$ and $m=2.$ Then~(26) takes
the form
\[
\ba{l}
\ds \frac{\partial {\bf v}}{\partial t}+v_k\frac{\partial {\bf v}}{
\partial x_k} =-g{\bf \hat e}+\frac1\rho \, \mbox{grad}\,p+\kappa \int_{{\bf
R}^3}\left\{ -\chi \left(\left\| {\bf y}\right\| ^2\right){\bf y}+\sum_{k=1}^2\frac
1{k!}\left[ \psi \left(\left\| {\bf y}\right\| ^2\right)\frac{\partial ^k{\bf v}}{
\partial {\bf x}^k}{\bf y}^k\right. \right.
\vspace{3mm}\\
\ds \qquad \left. \left. +\left \langle \frac{\partial ^k{\bf v}}{\partial {\bf x}
^k}{\bf y}^k,{\bf y}\right\rangle \left( \eta \left(\left\| {\bf y}
\right\| ^2\right)-\psi \left(\left\| {\bf y}\right\| ^2\right)\left\| {\bf y}\right\|
^{-2}\right) {\bf y}\right] \right\} d{\bf y},
\ea
\]
which upon integration using spherical coordinates simplif\/ies to
\begin{equation}
\frac{\partial {\bf v}}{\partial t}+v_k\frac{\partial {\bf v}}{
\partial x_k}=-g{\bf \hat e}+\frac 1\rho\,  \mbox{grad}\,p+\nu \Delta {\bf v}+\lambda
\,\mbox{grad}\,(\mbox{div}\,
{\bf v}),
\end{equation}
where
\[
\nu :=\frac{2\pi \kappa }{15}\int_0^\infty \left[ \eta (s ^2)s
^2-4\psi (s ^2)\right] s ^4ds
\]
and
\[
\lambda :=\frac{4\pi \kappa }{15}\int_0^\infty \left[ \eta (s
^2)s ^2-\psi (s ^2)\right] s ^4ds
\]
depend only on the density and the
particle-particle force functions described in Section~2
and may be assumed to be constants in many applications.

Observe that (27) is essentially just the momentum part of the Navier-Stokes
equations (cf.~\cite{blackmore:Pey,blackmore:Tab} and~\cite{blackmore:Tem}). There are at least two interesting
inferences that may be drawn from this result: Firstly, it provides a partial conf\/irmation of
the validity of our integro-partial dif\/ferential equation model as a predictive tool for
granular f\/low. Secondly, it lends support to the contention that the Navier-Stokes equations
are a good model for granular f\/low behavior obtained from simulation of the Newtonian
equations of motion.

\section{Flow through a tube: an exact solution}

In this section we obtain an exact solution of the approximate model~(27),~(23),~(24)
subject to appropriate boundary conditions, for the case of fully developed
(steady-state) granular f\/low, under the action of gravity, through a
vertical circular cylindrical pipe illustrated in Figure~3.
We assume that the density and pressure are constant, hence it suf\/f\/ices to
solve~(26) subject to some boundary conditions.

\strut\hfill

{\bf Figure 3:} Flow through a tube

\vspace{12cm}

Under the circumstances, it is convenient to recast (27) in terms
of standard cylindrical coordinates $(r,\theta ,z)$ with corresponding
velocity components $(u,v,w),$ where $u$ is the radial, $v$ the azimuthal
and $w$ is the vertical(axial) component of the f\/low velocity. The system
assumes the following form with respect to cylindrical coordinates:
\be
\ba{l}
\ds \frac{\partial u}{\partial t}+u\frac{\partial u}{\partial r}+\frac vr\frac{
\partial u}{\partial \theta }+w\frac{\partial u}{\partial z}-\frac{v^2}r
\vspace{3mm}\\
\ds \qquad =\nu \left[ \frac 1r\frac \partial {\partial r}\left( r\frac{\partial u}{
\partial r}\right) +\frac 1{r^2}\frac{\partial ^2u}{\partial \theta ^2}+
\frac{\partial ^2u}{\partial z^2}\right] +
\lambda \frac \partial {\partial r}\left[ \frac 1r\frac{\partial (ru)}{
\partial r}+\frac 1r\frac{\partial v}{\partial \theta }+\frac{\partial w}{
\partial z}\right],
\vspace{3mm}\\
\ds \frac{\partial v}{\partial t}+u\frac{\partial v}{\partial r}+\frac vr\frac{
\partial v}{\partial \theta }+w\frac{\partial v}{\partial z}+\frac{uv}r
\vspace{3mm}\\
\ds \qquad =\nu \left[ \frac 1r\frac \partial {\partial r}\left( r\frac{\partial v}{
\partial r}\right) +\frac 1{r^2}\frac{\partial ^2v}{\partial \theta ^2}+
\frac{\partial ^2v}{\partial z^2}\right] +
\frac \lambda r\frac \partial {\partial \theta }\left[ \frac 1r\frac{
\partial (ru)}{\partial r}+\frac 1r\frac{\partial v}{\partial \theta }+\frac{
\partial w}{\partial z}\right],
\vspace{3mm}\\
\ds \frac{\partial w}{\partial t}+u\frac{\partial w}{\partial r}+\frac vr\frac{
\partial w}{\partial \theta }+w\frac{\partial w}{\partial z}
\vspace{3mm}\\
\ds \qquad =-g+\nu
\left[ \frac 1r\frac \partial {\partial r}\left( r\frac{\partial w}{\partial
r}\right) +\frac 1{r^2}\frac{\partial ^2w}{\partial \theta ^2}+\frac{
\partial ^2w}{\partial z^2}\right] +
\lambda \frac \partial {\partial z}\left[ \frac 1r\frac{\partial (ru)}{
\partial r}+\frac 1r\frac{\partial v}{\partial \theta }+\frac{\partial w}{
\partial z}\right] ,
\ea \hspace{-5.45pt}
\ee
where $g$ is the acceleration of gravity.
We assume that the pipe has radius $R>0$ and that its length is so great
that the domain of the granular f\/low can be represented in idealized form as
\[
\Omega :=\left\{ (r,\theta ,z):0\leq r<R\right\} .
\]

Now we deal with the task of appending appropriate auxiliary data to~(28) on~$\Omega .$
As we are seeking a steady-state solution, we assume that the
velocity is independent of the time~$t.$ There remains the question of
realistic auxiliary data on the boundary $\partial \Omega .$ Of course,
$u\leq 0$ on $\partial \Omega $ is required by the geometry of the pipe
(assuming it is rigid and impenetrable). Over time, one may reasonably
expect the radial and azimuthal f\/luctuations in velocity along the inside
surface of the pipe to cease, so we shall assume that both~$u$ and~$v$
vanish on $\partial \Omega .$ As for the axial velocity along $\partial
\Omega $: the motion of a particle in contact with $\partial \Omega $ is
that of free fall with a resisting force due to friction. This suggests that
there is a constant limiting (or terminal) velocity along the wall of the
pipe (that is achieved in the long-term f\/low conf\/iguration), so it is
reasonable to assume that~$w$ is a negative constant along $\partial \Omega.$
In summary, we take the auxiliary data for (28) in $\Omega $ to be
\be
\ba{l}
\ds \frac{\partial u}{\partial t} =\frac{\partial v}{\partial t}=\frac{
\partial w}{\partial t}\equiv 0\qquad \mbox{in }\quad \Omega,
\vspace{3mm}\\
\ds u =v=0 \qquad \mbox{and }\quad w=-w_\infty \qquad \mbox{on }\quad
\partial \Omega,
\ea
\ee
where $w_\infty $ is a positive constant.

In view of the boundary conditions, it makes sense to seek a solution of
(28)--(29) with $u=v=0$ and $w=\varphi (r).$ Then the f\/irst two equations of~(27)
are trivially satisf\/ied and the third equation yields
\begin{equation}
\frac \nu r\frac d{dr}\left( r\frac{d\varphi }{dr}\right) =g.
\end{equation}
Integrating (30), we obtain
\[
\varphi =\frac g{4\nu }r^2+c_1\log r+c_2,
\]
where $c_1$ and $c_2$ are constants of integration. The solution should be
regular at $r=0,$ so we must set $c_1=0.$ Then~(29) leads to the following
solution:
\begin{equation}
u=v=0 \qquad \mbox{and}\qquad w=\frac g{4\nu }\left( r^2-R^2\right) -w_\infty .
\end{equation}
It is easy to check that (31) satisf\/ies (28)--(29).

\section{Inclined plane f\/low model}

The fully developed f\/low of particles down a two-dimensional inclined plane
will be studied in this section using the governing equations~(27),~(23),~(24).
As inclined plane f\/low has been extensively investigated (see, for example~[2]
and~[10]), we shall have an opportunity to compare the predictions based
upon the simple approximate model with the results obtained by other
researchers, thereby further testing the ef\/fectiveness of our approach.

Figure~4 depicts the f\/low geometry for a plane inclined at an angle
of~$\theta $ to the horizontal.

\strut\hfill

{\bf Figure 4:} Inclined plane flow

\vspace{11cm}

It is convenient to use a Cartesian coordinate system with $x$
measured down along the surface of the inclined plane and $y$-axis normal to
the plane and pointing into the f\/lowing layer of granular material. Here $u$
represents the component of the f\/low velocity along the $x$-axis and $h$ the
depth of the f\/lowing layer.
We assume that the density is constant and that the pressure gradient exactly
balances the gravitational force normal to the inclined plane throughout the f\/lowing
layer. Therefore, it suf\/f\/ices to solve~(27) along with the necessary boundary
conditions. Of course, the f\/igure embodies the usual
assumption that the granular f\/low is essentially two-dimensional.

A clockwise rotation of $\theta $ of the coordinate system and a
balancing of the gravitational and reaction forces normal to the inclined
plane yields the following pair of equations for the granular f\/low:
\begin{equation}
\frac{\partial u}{\partial t}+u\frac{\partial u}{\partial x}+v\frac{\partial
u}{\partial y}=g\sin \theta +\nu \left( \frac{\partial ^2u}{\partial x^2}+
\frac{\partial ^2u}{\partial y^2}\right) +\lambda \left( \frac{\partial ^2u}{
\partial x^2}+\frac{\partial ^2v}{\partial x\partial y}\right),
\end{equation}
\begin{equation}
\frac{\partial v}{\partial t}+u\frac{\partial v}{\partial x}+v\frac{\partial
v}{\partial y}=\nu \left( \frac{\partial ^2v}{\partial x^2}+\frac{\partial
^2v}{\partial y^2}\right) +\lambda \left( \frac{\partial ^2u}{\partial
x\partial y}+\frac{\partial ^2v}{\partial y^2}\right) ,
\end{equation}
where $v$ is the $y$-component of the velocity of the granular f\/low. Since
the f\/low is taken to be fully developed (steady-state), it is reasonable to
assume that both $u$ and $v$ are independent of the time $t.$ It is also
sensible to presuppose that the $y$-component of the velocity vanishes
identically and that $u$ is a function of $y$ only. With these assumptions~(33)
reduces to the trivial equation $0=0$ and we are left only with the
simple ordinary dif\/ferential equation
\begin{equation}
\frac{d^2u}{dy^2}=-\frac g\nu \sin \theta
\end{equation}
representing (32).

Appropriate auxiliary data for (34) are the free-boundary condition along
the free surface representing the interface between the f\/lowing particles
and the air that def\/ines the depth of the f\/lowing layer~$h$ as the smallest
number satisfying
\begin{equation}
\frac{du}{dy}(h)=0\qquad (h>0),
\end{equation}
and a slip condition along the inclined plane granular material interface
\begin{equation}
\frac{du}{dy}(0)=ku_\infty ,
\end{equation}
where $u_\infty $ is a limiting velocity along the surface of the plane. The
component $u_\infty $ may be the result of a partial balance between the
frictional properties of the plane and the particles and the gravitational
component of the force in the $x$-direction or a combination of
gravitational and frictional ef\/fects and some constant mass f\/low rate
supplied to the system. Here~$k$ is some nonnegative constant connected with
the nature of the shearing stress in the f\/lowing layer adjacent to the plane
that is related to the frictional characteristics of the plane and particles
and the dynamical state of the system.

Integrating (34) twice using (36), we obtain the solution
\begin{equation}
u=u(y)=-\left( \frac{g\sin \theta }{2\nu }\right) y^2+u_\infty \left(
ky+1\right) .
\end{equation}
Whence we determine the depth of the f\/lowing layer by substitution of (35)
in~(37); namely,
\begin{equation}
h=\frac{k\nu u_\infty }{g\sin \theta },
\end{equation}
for $\theta >0.$ A typical velocity prof\/ile is shown in Figure~5.

\strut\hfill

{\bf Figure 5:} Velocity profiles for inclined plane flow

\vspace{10cm}

The extremely simple nature of the solution (37) obtained from the governing
equation~(27) notwithstanding, it compares rather well with observations
from experimental studies and the predictions from more complicated f\/low
models (cf.~\cite{blackmore:And} and~\cite{blackmore:John}). For example, the form of the velocity
prof\/iles illustrated in Fig.~5 is qualitatively similar to those measured in
experiments and derived from more comprehensive constitutive equations.
Moreover, unlike some fairly popular models,~(38) shows that our approach predicts a
decrease in the depth of the f\/lowing layer with increasing inclination angle
of the plane~-- a property that is consistent with experimental observations.

\section{Vibrating bed model}

In this section we shall apply our model (27), (23), (24) to the study of
granular f\/lows in
a two-dimensional vibrating bed. Namely,
we consider the motion of a very large number of particles in a rectangular
container in the plane with f\/ixed vertical side walls and a horizontal bottom that is
oscillating periodically  in the vertical ($x_2$) direction. The only body force is a
gravitational force in the negative ($x_2$) direction and the interstitial and surrounding medium
is air which we assume has no ef\/fect on the granular f\/low.

At $t=0$ the particles are contained in the following region:
\be
K_0:=\{(x_1,x_2)\in{\bf R}^2: |x_1|<\sigma, \, x_2>0\},
\ee
where $\sigma >0$ is half of the width of the container. Then, the container
is subject to a vertical oscillation of the form $a\sin(\omega t)$ that is
illustrated in Figure~6.


\vfill

\pagebreak

{\bf Figure 6:} Granular material in a vibrating container

\vspace{12cm}

We shall use  boundary conditions at the walls similar to those
employed  in the previous section. Namely, we assume that the normal component of
the particle velocity near the wall is
equal to the normal component of the wall velocity. As for the tangential
direction, we use the equation~(36) in the following form
\be
\frac{\p v_T}{\p n}=-k v_T,
\ee
where $v_T$ denotes the relative tangential component of velocity between the
particle and the wall, $\p/\p n$ is the partial derivative in the outer normal direction.
The equation~(40) as well as~(36) represents a type of balance law between the
interparticle and particle-wall friction forces which has been used by other reseachers.
The (constant) coef\/f\/icient $k >0$ is a
measure of the boundary friction that we shall call the {\it wall friction
coefficient}. It would be most natural to use the particle size as the characteristic length
for the non-dimensionalization of~$k$, but it tends to zero in the continuum model limit.
Therefore we use for this purpose the space step of the numerical integration scheme.

In summary then, we take the following as the governing equations plus the
initial and boundary conditions for the granular f\/low in the planar vibrating bed:
\be
\ba{l}
\ds \frac {\p v_1}{\p t}+v_1\frac {\p v_1}{\p x_1}+v_2\frac {\p v_1}{\p
x_2}=\nu\left(\frac{\p^2v_1}{\p x_1^2}+
\frac {\p^2v_1}{\p x_2^2}\right)+\lambda\left(\frac {\p^2v_1}{\p x_1^2}+\frac {\p^2v_2}{\p
x_1\p x_2}\right),
\vspace{3mm}\\
\ds \frac{\p v_2}{\p t}+v_1\frac {\p v_2}{\p x_1}+v_2\frac {\p v_2}{\p
x_2}=\nu\left(\frac {\p^2v_2}{\p x_1^2}+
\frac {\p^2v_2}{\p x_2^2}\right)+\lambda\left(\frac {\p^2v_1}{\p x_1 \p
x_2}+\frac {\p^2v_2}{\p x_2^2}\right)
\ea
\ee
in $\Sigma:=\{(x,t): \,x\in\Omega_t, \, t>0\}$;
\be
v_1(x_1, x_2, 0)=v_1^0(x_1,x_2), \qquad v_2(x_1, x_2, 0)=v_2^0(x_1,x_2) \qquad
\mbox{at} \quad t=0
\ee
for all $x\in\Omega_0=\{x\in{\bf R}^2: \,|x_1|<\sigma, \,0<x_2<h\}$, where
the functions $v_1^0(x_1,x_2)$ and $v_2^0(x_1,x_2)$ determine the initial
velocity distribution;
\be
\frac {\p v_1}{\p x_2}=-kv_1, \qquad v_2=a\omega \cos(\omega t)
\ee
for all particles on the bottom, $\{(x_1, x_2):\, |x_1|<\sigma,
x_2=a\sin(\omega t)\}$, of the bed when $t>0$;
\be
v_1=0, \qquad \frac {\p v_2}{\p x_1}=-kv_2
\ee
for all particles on the left wall, $\{(x_1,x_2): \, x_1=-\sigma, \, x_2>a\sin(\omega t)\}$, of the
container when $t>0$;
\be
v_1=0, \qquad \frac {\p v_2}{\p x_1}=-kv_2
\ee
for all particles on the right wall, $\{(x_1,x_2): \, x_1=\sigma, \, x_2>a\sin(\omega t)\}$, of the
container when $t>0$; and
\be
\frac {\p {\bf v}}{\p n}=0
\ee
for all particles on the free-boundary, consisting of all points in
$\p\Omega_t\backslash\p K_t$, when $t>0$.

Now we shall consider  some numerical solutions to the model~(41) subject to
the initial and the boundary conditions~(42)--(46). To simplify our analysis, we shall ignore the
ef\/fects of surface waves and free-boundary components at the bottom of the container.

In order to avoid dif\/f\/iculties with vibrating bed boundary conditions like
$v_2(x, a\sin(\omega t))$ $=a\omega\cos(\omega t)$ at the bottom of the bed,
we shall write our equations in the vibrating system of coordinates:
\be
x=x^*, \qquad y=y^*+a\sin(\omega t), \qquad t=t^*.
\ee
The operators of partial dif\/ferentiation in this frame are
\be
\frac \p{\p x}=\frac \p{\p x^*}, \qquad \frac \p{\p y}=\frac \p{\p y^*}, \qquad \frac \p{\p
t}=\frac \p{\p t^*}-a\omega\cos(\omega t^*)
\frac \p{\p y^*}.
\ee
We have to include also into the system (41) the inertial force term
proportional to $a\omega^2 \sin(\omega t)$ and directed along the $y$-axis.
The governing system of equations in the "starred" system takes the following
form (we have dropped the index `$*$')
\be
\ba{l}
\ds v_{1,t}=\nu(v_{1,xx}+v_{1,yy})+a\omega\cos(\omega
t)v_{1,y}-\alpha(v_1v_{1,x}+v_2v_{1,y})+\lambda(v_{1,xx}+
v_{2,xy}),
\vspace{2mm}\\
\ds v_{2,t}=a\omega^2 sin(\omega t)+\nu(v_{2,xx}+v_{2,yy})+a\omega\, cos(\omega
t)v_{2,y}-\alpha(v_1v_{2,x}+v_2v_{2,y})
\vspace{2mm}\\
\ds \qquad +\lambda(v_{1,xy}+v_{2,yy}),
\ea
\ee
where $0\le x\le 2$, $0\le y\le 2$, and the boundary conditions can be written as
\be
\ba{l}
v_1(0,y,t)=0, \qquad v_1(2,y,t)=0, \qquad v_2(x,0,t)=0, \qquad v_2(x,2,t)=0,
\vspace{2mm}\\
\ds \frac {\p v_1}{\p y}(x,0,t)+kv_1(x,0,t)=0, \qquad \frac {\p v_1}{\p y}(x,2,t)=0,
\vspace{3mm}\\
\ds \frac {\p v_2}{\p x}(0,y,t)+kv_2(0,y,t)=0, \qquad
\frac {\p v_2}{\p x}(2,y,t)+kv_2(2,y,t)=0.
\ea
\ee

We use an explicit f\/inite dif\/ference scheme for solving the system (49)--(50)
with the spatial step $\Delta x=\Delta y=0.001$ and the time step
$\Delta t=10^{-5}$. Estimates show that such a small time step is needed to
satisfy the stability condition for the explicit f\/inite-dif\/ference scheme.

We investigated the system (49)--(50) using both multi-vortex and random initial
conditions. For certain ranges of the parameters (corresponding to the particle-particle
forces) the motion of the system starting with a multi-vortex conf\/iguration changes
rapidly into a pair of vortices that persists for a long time (relative to the period of the
forced oscillations). The centers of this ``stable'' vortex pair oscillate
with small amplitude synchronistically with the  forced oscillations.
In some cases this vortex pair evolved into a single vortex over a very large
time period. Increase of the constant $\nu$ results in a corresponding  increase of  the
particle-particle friction and leads to  damping of the vorticity (see
Figure~7).

When the motion starts from a random initial velocity distribution we also
observed the ``stable'' vortex type of motion and
bifurcation between  dif\/ferent types of relatively stable patterns (Figure~8).
Some of the values used for the control parameters were
$\omega=2$, $a=1$, $\nu=0.3$, $\lambda=1.0$
in the case of the vortex type of motion, and
$\omega=3$, $a=1$, $\nu=1.0$, $\lambda=0.5$
in the case of a mixing motion. These types of  particle
dynamics are in agreement with  experimental observations and  computer
simulation results~\cite{blackmore:Lan,blackmore:Poesch}.
We note also that Hayakawa and Hong~\cite{blackmore:Hay} obtained similar results from
numerical solutions of their models but they assumed no-slip boundary conditions
at the walls which are not physically realistic for vibrating bed granular f\/lows.

Similar types of the f\/low behavior were obtained by Bourzutschky and
Miller~\cite{blackmore:Bourz}
for their Navier-Stokes models. By using negative slip boundary conditions in numerical
experiments corresponding to granular f\/lows with a high mobility boundary layer,
they obtained experimentally observable vortex type solutions.
Unlike us, they did not obtain convective f\/low behavior coinciding with
experimentally observed results for possible values of the wall friction coef\/f\/icient. A possible
explanation for this discrepancy between their f\/indings and ours may be the fact that we included
the gravity force directly in our model and they did not.

As mentioned above, we have suppressed the free-boundary conditions that occur
in an actual vibrating bed in our numerical experiments. This has been done to simplify
the numerical solution of the problem, since incorporation of the free-boundaries
signif\/icantly complicates the problem and for us is still
in the developmental stage. Preliminary results indicate that the addition of
free-boundaries will still result in the appearance of ``stable'' convective vortices,
and we plan to demonstrate this in a forthcoming
paper. Apparently, the frictional ef\/fects of the walls is the primary mechanism
in the generation of convective rolls. For now then, our analysis of the vortices  must be
considered to be of a local rather than a global nature.

\strut\hfill

{\bf Figure 7:} Motion of particles starting with
four-vortex initial configuration (numerical solutions)

\vfill

\pagebreak

\strut\hfill

{\bf Figure 8:} Motion of particles starting with random
initial velocity distribution (numerical solutions)

\vfill

\pagebreak

\section{Concluding Remarks}

Starting with well-established representations for particle-particle normal
and tangential frictional forces (based on sound theoretical principles and
a large body of experimental observations), we derived a new class of
integro-partial dif\/ferential equations to describe the velocity f\/ield in the
granular f\/low of rough, inelastic particles. These granular f\/low models were
obtained by taking a dynamical limit as $N\rightarrow \infty $ of the
Newtonian system of dif\/ferential equations of motion of an $N$-particle
array using integral averages of an assumed uniform distribution of
particles comprising the f\/low f\/ield. Then by employing Taylor series
expansions of key variables of the f\/low f\/ield, we were able to obtain an
inf\/inite collection of approximations of the model equations in the form of
a system of three nonlinear partial dif\/ferential equations for the velocity
components of the granular f\/low. The simplest of these approximate models,
obtained by retaining only the f\/irst two terms in the Taylor expansions, is
a system of equations that is signif\/icantly less complicated than most of
the continuum models currently being used to investigate particle f\/low
dynamics.

Our models, and especially the simplest of the approximations, certainly do
not incorporate as much of the physics involved in granular f\/lows as do the
more comprehensive partial dif\/ferential equation models, yet they appear to
be quite promising instrumentalities for the prediction of particle f\/low
behavior. A good indication of this is the results of our application of the
simplest model to granular f\/low through a vertical tube,  fully developed
f\/low down an inclined plane and f\/low in a vibrating bed
which produced relatively simple solutions that
compared remarkably well, in a qualitative sense, with experimental
observations and the predictions of more complete models. This suggests
that, in spite of the simplifying assumptions we used in the derivation,
these models may be capable of accurately predicting dynamical properties of
a wider range of granular f\/low conf\/igurations than one might imagine. And
that they certainly warrant further investigation and testing. Moreover, the
new models are far more amenable to analysis (particularly from the
viewpoint of inf\/inite-dimensional dynamical systems theory) than the
majority of governing equations in the literature. Therefore it is quite
possible that one may be able to apprehend important new insights into
several elusive granular f\/low phenomena from a more penetrating mathematical
investigation of the properties of the new equations.

In the near future we plan to undertake an intensive analytical and
computational study of the models introduced in this paper. For example, we
shall use dynamical systems theory to identify and analyze such phenomena as
inertial manifolds, bifurcations, strange attractors and regimes of
spatio-temporal chaos that will then be correlated to a variety of complex
granular f\/low behaviors. In addition, it should be useful to develop and
implement algorithms for the approximate numerical solution of the models,
and then compare the results obtained with those from
simulations,experimental studies and other governing equations. We shall
begin this program of investigation by conducting a more thorough
analysis of vibrating bed f\/lows and also studying granular f\/low
in hoppers.

\subsection*{Acknowledgments}
This work was partially supported by DOE Contract DE-FG22-95PC95203, NSF
Grant EEC-9420597 and a grant from the New Jersey Commission on Science and
Technology. The authors are indebted to a referee whose insightful comments
led to an improvement of the original version of this paper.

\label{blackmore-lp}


\begin{thebibliography}{99}

\footnotesize


\bibitem{blackmore:An} An L. and Pierce A., A Weakly Nonlinear Analysis of
Elasto-Plastic Microstructure, {\it SIAM J. Appl. Math.}, 1995, V.55, 136--155.

\bibitem{blackmore:And} Anderson K. and Jackson R., A Comparison of Some Proposed
Equations  of Motions of Granular Materials for Fully Developed Flow Down Inclined
Planes,  {\it J. Fluid Mech.}, 1992, V.241, 145--168.

\bibitem{blackmore:Baxter} Baxter G.W. and Behringer R.P., Cellular Automata Models of
Granular Flow, {\it Phys. Rev. A}, 1990, V.42, 1017--1020.

\bibitem{blackmore:Bourz} Bourzutschky M. and Miller J., ``Granular'' Convection in a
Vibrated Fluid, {\it Phys. Rev. Lett.}, 1995, V.74, 2216--2219.

\bibitem{blackmore:Caram} Caram H. and Hong D.C., Random-Walk Approach to Granular Flows,
{\it Phys. Rev. Lett.}, 1991, V.67, 828--831.

\bibitem{blackmore:Fan} Fan L. and Zhu C.,  Principles of Gas-Solid Flows,
Cambridge Univ. Press (manuscript).

\bibitem{blackmore:Far} Farrel M., Lun C. and Savage S., A Simple Kinetic Theory for
Granular Flow of Binary Mixtures of Smooth, Inelastic Spherical Particles, {\it Acta
Mech.}, 1986, V.63, 45--60.

\bibitem{blackmore:Gard} Gardiner C. and Schaef\/fer D., Numerical Simulation of
Uniaxial Compression of Granular Material with Wall Friction,  {\it SIAM J. Appl. Math.},
1994, V.54, 1676--1692.

\bibitem{blackmore:Goldh} Goldhirsch I., Tan M.-L. and Zanetti G., A Molecular Dynamical
Stady of Granular Fluids I: the Unforces Granular Gas in Two Dimensions,
{\it J. Sci. Comp.}, 1993, V.8, 1--40.

\bibitem{blackmore:Gold} Goldshtein A. and Shapiro M., Mechanics of Collisional Motion of
Granular Materials, Part 1. General
Hydrodynamic Equations, {\it J. Fluid Mech.}, 1995, V.282, 75--114.

\bibitem{blackmore:Hay} Hayakawa H. and Hong D., Two Hydrodynamical Models  of Granular
Convection, in Powders and Grains 97,
Benringer and Jenkins (eds), Balkema, Rotterdam, 1997, 417--420.

\bibitem{blackmore:Jeni} Jenike A. and Shield R., On the Plastic Flow of Coulomb
Solids Beyond Original Failure, {\it J. Appl. Mech.}, 1959, V.26, 599--602.

\bibitem{blackmore:Jen} Jenkins J. and Savage S., A Theory for Rapid Flow of Identical
Smooth, Nearly Elastic Particles, {\it J.~Fluid Mech.}, 1983, V.130, 187--202.

\bibitem{blackmore:Jen2} Jenkins J. and Richman M., Grad's 13-moment System for a
Dense Gas of Inelastic Spheres,  {\it Arch. Rat. Mech. Anal.}, 1985, V.87,
355--377.

\bibitem{blackmore:John} Johnson P., Nott P. and Jackson R., Frictional-Collisional
Equations of Motion for Particulate Flows and Their Application to Chutes, {\it J. Fluid
Mech.}, 1990, V.210, 501--535.

\bibitem{blackmore:Lan2} Lan Y. and Rosato A., Macroscopic Behavior of a Vibrating
Bed of Smooth Inelastic  Particles, {\it Phys. Fluids}, 1995, V.7,  1818--1831.

\bibitem{blackmore:Lan} Lan Y. and Rosato A., Convection Related Phenomena in Vibrating
Granular Beds, {\it Phys. Fluids}, 1997, V.9, N~12, 3615--3624.

\bibitem{blackmore:Lun} Lun C., A Kinetic Theory for Granular Flow of Dense, Slightly
Inelastic, Slightly Rough Spheres, {\it J.~Fluid Mech.}, 1991, V.233,  539--559.

\bibitem{blackmore:Lun3} Lun C. and Savage S., A Simple Kinetic Theory for Granular
Flow of Rough, Inelastic Spherical Particles, {\it J. Appl. Mech.}, 1987, V.54,
47--53.

\bibitem{blackmore:McN} McNamara S. and Luding S, Energy Flows in Vibrated Bed Granular
Media, {\it Phys. Rev. E}, 1998, V.58, 813.

\bibitem{blackmore:Num} Numan K. and Keller J., Ef\/fective Viscosity of a Periodic
Suspension, {\it J. Fluid  Mech.}, 1984, V.142, 269--287.

\bibitem{blackmore:Pasq} Pasquarell G., Granular Flows: Boundary Conditions for
Slightly Bumpy Walls, {\it  ASCE J. Eng. Mech.}, 1991, V.117, 312--318.

\bibitem{blackmore:Pey} Peyret R. and Taylor T.,  Computational Methods for
Fluid Flow, Springer-Verlag, New York, 1983.

\bibitem{blackmore:Pit} Pitman E.B., Gudonov Method for Localization in Elastoplastic Granular Flow,
{\it Int. J.~Num. Anal. Methods}, 1993, V.17, 385--400.

\bibitem{blackmore:Poesch} P{\"o}eschel T. and Herrmann H.J., Size Segregation and
Convection, {\it Europhys. Lett.}, 1995, V.29, 123--128.

\bibitem{blackmore:Raj} Rajagopal K., Existence of Solutions to the Equations
Governing the Flow of  Granular Material, {\it Euro. J. Mech. B~/~Fluids}, 1992, V.11,
265--276.

\bibitem{blackmore:Rich} Richman M., Boundary Conditions for Granular Flows at
Randomly Fluctuating Bumpy Boundaries, in  Advances in Micromechanics of Granular
Materials, Shen H. et al. eds., Elsevier, Amsterdam, 1992, 111--122.

\bibitem{blackmore:Ros4} Rosato A., Dave R., LaRosa A. and Mosch E., Experimental
Study of Vibrational Size Segregation, in Proc. First Int. Particle Tech. Forum, AIChE,
1994, 325--330.

\bibitem{blackmore:Sav} Savage S., The Mechanics of Rapid Granular Flows,
{\it Adv. in Appl. Mech.}, 1984, V.24, 289--366.

\bibitem{blackmore:Sav2} Savage S., Studies of Granular Shear Flow. Wall Slip
Velocities, ``Layering'' and Self-Dif\/fusions, {\it Mech. Gran. Mater.}, 1993, V.16,
225--238.

\bibitem{blackmore:Sav3} Savage S. and Jef\/frey D., The Stress Tensor in a Granular
Flow at High Shear Rates, {\it J. Fluid Mech.}, 1981, V.110,  225--272.

\bibitem{blackmore:Scha} Schaef\/fer D., Instability in the Evolution Equations
Describing Incompressible Granular Flow, {\it J.~Diff. Eq.}, 1987,
V.66, 19--50.

\bibitem{blackmore:Scha2} Schaef\/fer D., Shearer M. and Pitman, E., Instability in
Critical State Theories of Granular Flow, {\it SIAM J. Appl. Math.}, 1990, V.50, 33--47.

\bibitem{blackmore:Shen} Shen H. and Ackermann N., Constitutive Equations for a
Simple Shear Flow of a Disk Shaped Granular Mixture, {\it Int. J.~Eng. Sci.}, 1984, V.7, 829--840.

\bibitem{blackmore:Swin} Swinney H., Umbanhowar P. and Melo F., Stripes, Squares,
Hexagons and Localized Structures in Vertically Vibrated Granular Layers, in:
Powders and Grains 97, Behringer and Jenkins (eds), 369--372, 1997, Balkema,
Rotterdam, SBN 90 5410 8843.

\bibitem{blackmore:Tab} Tabor M.,  Chaos and Integrability in Nonlinear
Dynamics, Wiley, New York, 1989.

\bibitem{blackmore:Tem} Temam R.,  Inf\/inite Dimensional Dynamical Systems in Mechanics
and Physics, Springer-Verlag, New York, 1988.

\bibitem{blackmore:Tsim} Tsimring L. and Aranson I., Localized and Cellular Patterns in a
Vibrated Granular Layer, {\it Phys. Rev. Lett.}, 1997, V.79, 213--216.

\bibitem{blackmore:Walt} Walton O., Numerical Simulation of Inelastic, Frictional
Particle-Particle Interactions, in Particulate Two-phase Flow, edited by M.C.~Roco,
Butterworth-Heinemann, Boston, 1992, 884--911.
\end{thebibliography}
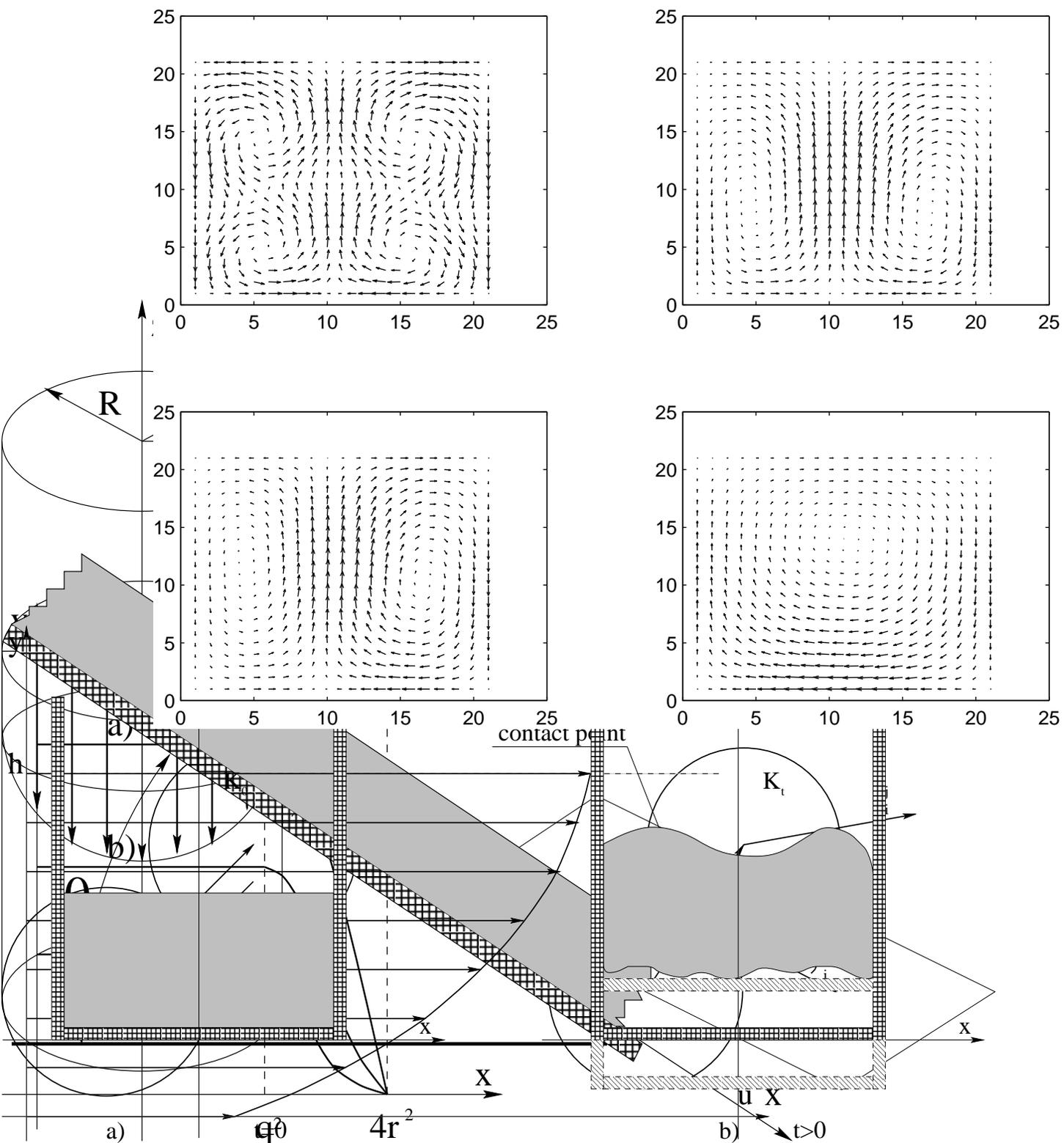
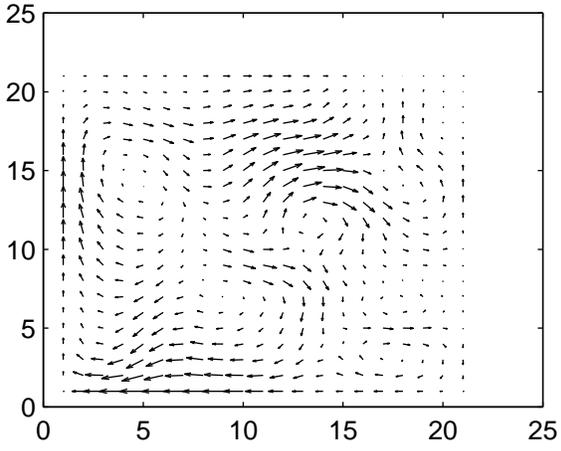
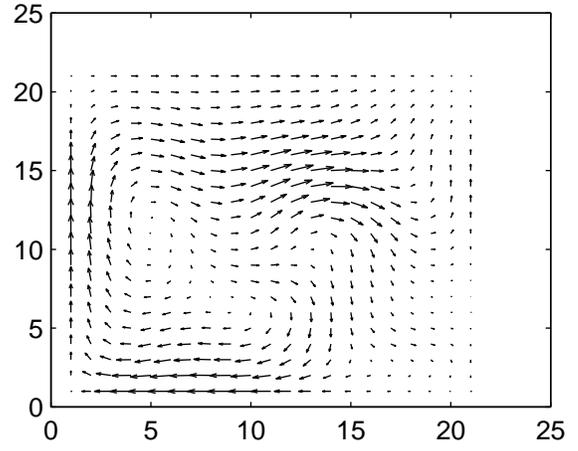
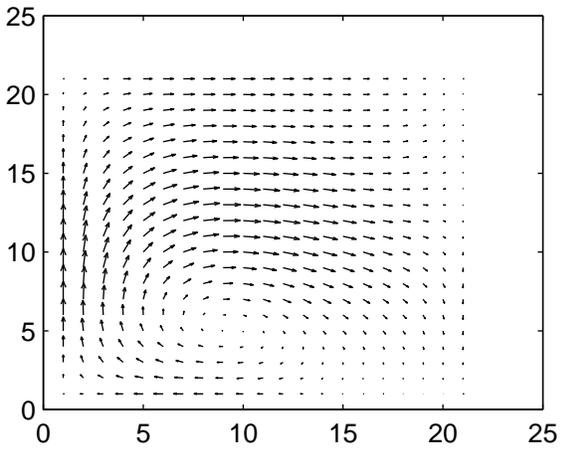
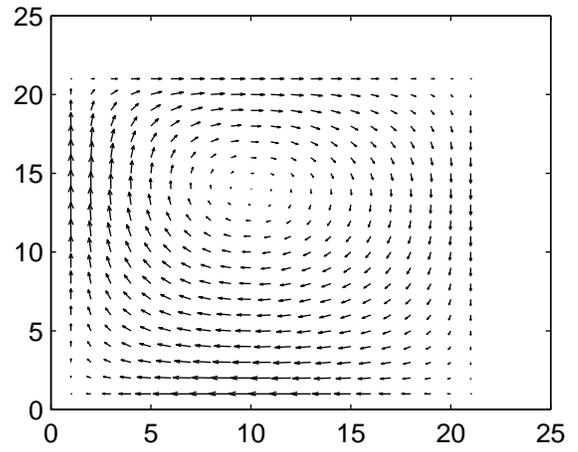
\end{document}